\documentclass[%
 reprint,
superscriptaddress,
 amsmath,amssymb,
 aps,
 prl,
 longbibliography
]{revtex4-1}

\newcommand{\NM}{$\sum m_{\nu}$}

\usepackage{lineno,xcolor}
\usepackage{amsmath}
\usepackage{url}
\usepackage{graphicx}
\usepackage{dcolumn}
\usepackage{bm}
\usepackage{subfigure}

\newcommand\mnras{Mon. Not. R. Astron. Soc.} 
\newcommand\jcap{J.~Cosmol. Astropart. Phys.} 
\newcommand\aap{A\&A}                

\usepackage[most]{tcolorbox}

\begin{document}

\preprint{APS/123-QED}

\title{Upper Bound of Neutrino Masses from Combined Cosmological Observations and Particle Physics Experiments}

\author{Arthur Loureiro}
 \email{arthur.loureiro.14@ucl.ac.uk}
  \affiliation{Department of Physics and Astronomy, University College London, Gower Street, London WC1E 6BT, United Kingdom}%
\author{Andrei Cuceu}%
 \email{andrei.cuceu.14@ucl.ac.uk}
\affiliation{Department of Physics and Astronomy, University College London, Gower Street, London WC1E 6BT, United Kingdom}%

\author{Filipe B. Abdalla}
\affiliation{Department of Physics and Astronomy, University College London, Gower Street, London WC1E 6BT, United Kingdom}
\affiliation{Department of Physics and Electronics, Rhodes University, P.O. Box 94, Grahamstown, 6140, South Africa}

\author{Bruno Moraes}
\affiliation{Department of Physics and Astronomy, University College London, Gower Street, London WC1E 6BT, United Kingdom}
\affiliation{Instituto de Fisica, Universidade Federal do Rio de Janeiro, 21941-972, Rio de Janeiro, Brazil}

\author{Lorne Whiteway}
\affiliation{Department of Physics and Astronomy, University College London, Gower Street, London WC1E 6BT, United Kingdom}

\author{Michael McLeod}
\affiliation{Department of Physics and Astronomy, University College London, Gower Street, London WC1E 6BT, United Kingdom}

\author{Sreekumar T. Balan}
\affiliation{Department of Physics and Astronomy, University College London, Gower Street, London WC1E 6BT, United Kingdom}

\author{Ofer Lahav}
\affiliation{Department of Physics and Astronomy, University College London, Gower Street, London WC1E 6BT, United Kingdom}

\author{Aur\'elien Benoit-L\'evy}
\affiliation{CNRS, UMR 7095, Institut d'Astrophysique de Paris, F-75014, Paris, France}%

\author{Marc Manera}
\affiliation{Institut de F\'isica d'Altes Energies, The Barcelona Institute of Science and Technology, Campus UAB, 08193 Bellaterra (Barcelona), Spain}%
\affiliation{Kavli Institute for Cosmology, University of Cambridge, Madingley Road, Cambridge CB3 0HA, United Kingdom}

\author{Richard P. Rollins}
\affiliation{Jodrell Bank Centre for Astrophysics, University of Manchester, Oxford Road, Manchester M13 9PL, United Kingdom}%

\author{Henrique S. Xavier}
\affiliation{Instituto de Astronomia, Geof\'isica e Ci\^encias Atmosf\'ericas, Universidade de S\~ao Paulo, Rua do Mat\~ao, S\~ao Paulo 05508-090, Brazil }%

\date{\today}

\begin{abstract}
We investigate the impact of prior models on the upper bound of the sum of neutrino masses, \NM{}. Using data from large scale structure of galaxies, cosmic microwave background, type Ia supernovae, and big bang nucleosynthesis, we argue that cosmological neutrino mass and hierarchy determination should be pursued using exact models, since approximations might lead to incorrect and nonphysical bounds. We compare constraints from physically motivated neutrino mass models (i.e., ones  respecting oscillation experiments) to those from models using standard cosmological approximations. The former give a consistent upper bound of $\sum m_{\nu} \lesssim 0.26$ eV (95\% CI) and yield the first approximation-independent upper bound for the lightest neutrino mass species, $m_0^{\nu} < 0.086$ eV (95\% CI). By contrast, one of the approximations, which is inconsistent with the known lower bounds from oscillation experiments, yields an upper bound of $\sum m_{\nu} \lesssim 0.15$ eV (95\% CI); this differs substantially from the physically motivated upper bound.


\end{abstract}

\pacs{Valid PACS appear here}
\maketitle

\textit{Introduction.}--Particle physics experiments in the late 1990s, such as Super-Kamiokande \citep{Kamiokande1998}, and recent experiments, such as SNO \citep{2002SNO}, KamLAND \citep{2005KamLAND}, and others \citep{2008MINOS,2012RENOExperiment,AbeNeutrino2014}, have established the existence of massive neutrinos, taking a first step beyond the standard model of particle physics. 
Current global fits to data from several neutrino oscillation experiments obtained constraints for two different mass squared splittings: from solar neutrino experiments, $\Delta m_{21}^2 \equiv m_2^2 - m_1^2 \approx 7.49^{+0.19}_{-0.17} \times 10^{-5}$ eV$^2$, and from atmospheric neutrinos, $|\Delta m_{31}^2| \equiv |m^2_3 - m_1^2| \approx 2.484^{+0.045}_{-0.048} \times 10^{-3}$ eV$^2$ (1$\sigma$ uncertainties) \citep{2014Gonzalez-GarciaNeutrino}. These measurements imply that at least two of the neutrino mass eigenstates are nonzero and, given that the sign of $\Delta m_{31}^2$ is unknown, that two scenarios are possible, related to the ordering of the masses: $m_1 < m_2 \ll m_3$, known as the \textit{normal hierarchy} (NH), or $m_3 \ll m_1 < m_2$, the \textit{inverted hierarchy} (IH). Current neutrino experiments will not be able to break the degeneracy between these two hierarchies (or orderings) in the near future \citep{2014Blennow}. However, by considering the lightest neutrino mass eigenstate to be zero we see that these experiments set a lower bound for the sum of neutrino masses,  $\sum m_{\nu} \equiv \sum_{i=1}^3 m_{\nu, i}$, as follows:  $\sum m_{\nu}^{\text{NH}} > 0.0585 \pm 0.00048$ eV or $\sum m_{\nu}^{\text{IH}} > 0.0986 \pm 0.00085$ eV \citep{2016JCAP...11..035H,2018UpdateNeutrinoMass,2018LongNeutrinoMassPior,2018Gariazzo}.

\begin{table*}
  \centering
  \caption{Neutrino mass models considered in this work, the neutrino parameters sampled, and the 95\% CI upper bounds on both \NM{} and $m_{0}^{\nu}$. Results were obtained using a combination of BOSS large scale structure $C_{\ell}$'s, Planck CMB temperature and polarization, Planck lensing, type Ia supernovae from Pantheon, and BBN measurements of D/H data. Models 1--4 also include constraints from oscillation experiments. }
  \label{Tb:Models}
  \begin{tabular}{cp{85mm}|p{33mm}|c|c}
    \hline
    \hline
    & Model description & \centering$\nu$-parameters & $\sum m_{\nu}$ & $m_0^{\nu}$\\
    & & & \small{[95\% CI]} & \small{[95\% CI]}\\[0.1cm]
    \hline
    
     1 & Both hierarchies, $m_0^{\nu}$ parametrization, sampling $|\Delta m_{31}^2|$ and $\Delta m_{21}^2$ from Gaussian priors & \centering $m_0^{\nu}$, $\mathcal{H}$, $|\Delta m_{31}^2|$, $\Delta m_{21}^2$   & $< 0.264$ eV & $< 0.081$ eV \\
     
     2 & Both hierarchies,  $m_0^{\nu}$ parametrization with $|\Delta m_{31}^2|$ and $\Delta m_{21}^2$ fixed to their central value & \centering $m_0^{\nu}$, $\mathcal{H}$ & $<0.275$ eV & $< 0.086$ eV\\
     
     3 & NH,  $m_0^{\nu}$ parametrization, fixed mass splittings & \centering $m_0^{\nu}$  & $< 0.261$ eV & $< 0.085$ eV \\
     
     4 & IH,  $m_0^{\nu}$ parametrization, fixed mass splittings & \centering $m_0^{\nu}$  & $< 0.256$ eV & $< 0.078$ eV\\
     
     \hline
     5 & NH approximation, $N_{\nu}=1$  & \centering $\sum m_{\nu}$  & $< 0.154$ eV & ...\\
     
     6 & IH approximation, $N_{\nu}=2$ & \centering $\sum m_{\nu}$  & $< 0.215$ eV & ...\\
     
     7 & Degenerated masses approximation, $N_{\nu}=3$ & \centering $\sum m_{\nu}$  & $< 0.270$ eV & ...\\
     \hline
     8 & No massive neutrinos, i.e., $N_{\nu}=0$ &  \centering ...  & ... & ...\\
     9 & Fixed to NH's lower bound, $\sum m_{\nu} = 0.06$ eV &  \centering ...  & ... & ...\\
     \hline
     \hline
  \end{tabular}
\end{table*}

From a different perspective, cosmological surveys have the potential to probe the sum of neutrino masses \citep{2007FBA,2002-OferNeutrinos,Thomas2010Neutr,2018HeavensNeutrino}, and also to constrain the neutrino mass hierarchy \citep{2003HannestadNeutrino,2016JCAP...11..035H,2018HeavensNeutrino}. The large scale structure of galaxies in the Universe is sensitive to the sum of neutrino masses and the number of massive neutrino species, $N_{\nu}$, since the cosmic energy density ratio for massive neutrinos in a $\Lambda$CDM model is
\begin{equation}
    \Omega_{\nu} = \sum_i^{N_{\nu}}\left[\left(\frac{G}{\pi^2H_0^2}\right)\int d^3p_i \frac{\sqrt[]{p_i^2 + m_{\nu,i}^2}}{(e^{p_i/T_{\nu,i}} + 1)} \right].
\end{equation} 

\noindent For the case of degenerate masses and after neutrinos start behaving nonrelativistically, this can be approximated by $ \Omega_{\nu} \approx \sum m_{\nu}/(92.5\, h^2\text{ eV})$ \citep{Thomas2010Neutr}. This last approximation is at the core of the approach taken by most cosmological analyses when probing the related neutrino parameters; this leads to  95\% CI upper bounds on \NM{} as low as $< 0.12$ eV from Ly-$\alpha$ measurements \citep{2015LyAlpha-Deg} and also from the latest Planck Collaboration results \cite{PlanckCosmology2018}. A complete review of neutrino mass ordering in cosmology and particle physics can be found in Refs. \cite{2012Julien-Deg,2018MassOrderingReview}.

In this Letter, we investigate the impact of different classes of neutrino mass modeling strategies on cosmological parameters and neutrino constraints. This test is performed with the latest cosmological data, namely a tomographic analysis in harmonic space applied to the largest spectroscopic galaxy sample to date, the BOSS DR12 \citep{2018LoureiroBOSS}, combined with Planck cosmic microwave background (CMB) temperature, polarization, and lensing \citep{2016PlanckCosmology}, Pantheon supernovae compilation data \cite{2018Pantheon}, BBN measurements of the deuterium-hydrogen fraction \citep{2018BBN-Measurements}, and, in some of the models, the latest neutrino mass squared splitting constraints from particle physics \cite{2014Gonzalez-GarciaNeutrino}.

\textit{Neutrino mass models.}--We compare the impact of seven different neutrino model priors on the upper bound of \NM{}. Each of the models in this section is probed together with all other $\Lambda$CDM parameters and $N_{\text{eff}}$, the effective number of relativistic species, and datasets are combined at the likelihood level--details are given in the subsequent sections. These prior models are subdivided into two categories: exact models and cosmological approximations. The exact models incorporate particle physics constraints from neutrino oscillation experiments via modeling \NM{}, using a parametrization based on the smallest neutrino mass $m_0^{\nu}$ \cite{2012Hannestad,2016Hannestad,2018HeavensNeutrino}. For the normal hierarchy, we have
\begin{align}
    \sum m_{\nu}^{NH} =\,  m^{\nu}_{0} & + \sqrt[]{\Delta m_{21}^2 + (m^{\nu}_{0})^2} \nonumber \\
    & + \sqrt[]{|\Delta m_{31}^2| + (m^{\nu}_{0})^2}
\end{align} 

\noindent while in the inverted hierarchy 
\begin{align}
    \sum m_{\nu}^{IH} =\,  m_0^{\nu} & + \sqrt[]{|\Delta m_{31}^2| + (m_0^{\nu})^2} \nonumber\\
    & + \sqrt[]{ |\Delta m_{31}^2| + \Delta m_{21}^2 + (m_0^{\nu})^2}.
\end{align} 
In what follows these will be referred to as the $m_0^{\nu}$ parametrization.

More explicitly, we use four exact models. {Model 1} samples a binary switch parameter $\mathcal{H}$, allowing the analysis to change between two hierarchies with the same prior volume, while also sampling the particle physics constraints for the mass splittings, $\Delta m_{21}^2$ and $|\Delta m_{31}^2|$, from Gaussian priors incorporating the errors in these measurements. {Model 2} is similar to model 1 but fixes the particle physics constraints to their central values: $\Delta m_{21}^2 = 7.49\times 10^{-5}$ eV$^2$ and $|\Delta m_{31}^2| = 2.484\times 10^{-3}$ eV$^2$. {Model 3} (respectively {model 4}) fixes the mass splittings to their central values while also fixing the hierarchy to be normal (respectively inverted).

The second class of models, the cosmological approximations, are related to degenerated scenarios in which $\sum m_{\nu} = N_{\nu}\times m_{\text{eff}}$, where $m_{\text{eff}}$ is an effective mass, equal for each massive neutrino species. For each of these models, $N_{\nu}$ is fixed to a specific value and \NM{} is sampled. {Model 5} is a NH approximation with $N_{\nu} = 1$; i.e., we approximate the two lower mass neutrino species to $m_1 = m_2 = 0$ \citep{2003HannestadNeutrino,2014Battye-Deg-1Mass,2016Giusarma-Deg-InvApp-NormAppr,2016PlanckCosmology,2018LoureiroBOSS}. Next, in a similar way, {model 6} is an IH approximation, where the lightest neutrino species is considered to be massless, which implies that $N_{\nu} = 2$ \citep{2012Hannestad,2016Giusarma-Deg-InvApp-NormAppr}. The last model in this class, {model 7}, is the most commonly used in standard cosmological analysis: the degenerate neutrino mass spectrum case, where $N_{\nu} = 3$ and $\sum m_{\nu} = 3m_{\text{eff}}$ \citep{2012Julien-Deg,2013Giusarma-Deg,2014Battye-Deg-1Mass,2015LyAlpha-Deg,2016Cuesta-Deg,2016BOSSCosmology,2016Giusarma-Deg-InvApp-NormAppr,2017Achidiacono-Deg,2017Cuchout-DegCase,2018PlanckCosmology,2017Vagnozzi-3deg,2018UpdateNeutrinoMass}. 

We also compare these seven models to cases where the \NM{} parameter is fixed to the most common values found in the literature for $\Lambda$CDM analysis \cite{2017arXiv170801530D,2017MNRAS.465.1454H,PlanckCosmology2018,2016BOSSCosmology}. {Model 8} assumes no massive neutrinos, while {model 9} fixes it to the minimum possible value for the NH, $\sum m_{\nu} = 0.06$ eV, and sets $N_{\nu} = 3$ (as in the $\Lambda$CDM approach taken by the Planck Collaboration \cite{2016PlanckCosmology,PlanckResults2015}). Since our analysis uses a nested sampler \citep{2008FerozHobson,PlinyRichardThesis}, Bayesian evidences are calculated for each model and combination of datasets. One can then use the ratio of Bayesian evidences between different models (the Bayes factor) to quantify which model is preferred by the datasets. Models 8 and 9 were added to the list of models with the purpose of checking this ratio.

A summary of each model, together with the relevant neutrino mass parameters sampled and the upper bounds for \NM{} and $m_{0}^{\nu}$ at 95\% credible interval (CI), can be found in Table \ref{Tb:Models}.



\textit{Assumptions.}--Since the newest analysis from the Planck Collaboration demonstrates that the Universe is flat to within 0.2\% precision, in this analysis we assume a flat $\Lambda$CDM scenario with massive neutrinos. The equation of state of dark energy is fixed to the cosmological constant case $w=-1$. We also assume the possibility of extra effective ultrarelativistic particles, which are probed via the $N_{ur}$ parameter--this parameter is degenerate with the decoupling of massive neutrinos at different temperatures, and, for simplicity, we assume the same decoupling temperature. As the galaxy clustering information comes from BOSS DR12 angular power spectra, no fiducial cosmology was assumed for this sample (as explained in Ref. \citep{2018LoureiroBOSS}). Priors for the standard $\Lambda$CDM parameters and nuisance parameters are as described in Table 3 in Ref. \citep{2018LoureiroBOSS}. The neutrino related priors are $\sum m_{\nu} \in [0.0, 1.0]$ eV, $m_{0}^{\nu} \in [1\times 10^{-4},0.3]$ eV, and $N_{ur} \in [-N_{\nu},(6-N_{\nu})]$ for extra ultrarelativistic species or the temperature neutrinos decouple \cite{2012Julien-Deg,2018Gariazzo}. This $N_{ur}$ dependency on $N_{\nu}$ for the extra ultrarelativistic species prior ensures an equivalent $N_{\text{eff}}$ prior on all models as $N_{\text{eff}}$ is a derived parameter in our analysis. Since models 5 and 6 do not have $N_{\nu} = 3$, we varied $N_{\text{eff}}$ to ensure that having $N_{\nu}$ different than $3$ does not have an impact on any other neutrino parameters. For models sampling the hierarchy parameter $\mathcal{H}$, the prior assigns equal odds for both hierarchies.


\textit{Data and methodology.}--Our main galaxy sample is a modified version of the BOSS DR12 large scale structure sample from Ref. \citep{BOSSCatalogue2016} as presented in Ref. \citep{2018LoureiroBOSS}. This sample is divided into 13 tomographic bins of $\Delta z = 0.05$ in a redshift range of $0.15 < z < 0.80$ containing a total of $\sim 1.15$M spectroscopic galaxies over more than 9000 deg$^2$ in the sky. Angular power spectra of these galaxies are measured using a pseudo-$C_{\ell}$ estimator (PCL) \citep{Thomas2011,Peebles1973,Efstat2004,2018-FreeCitation} in a bandwidth of $\Delta\ell =8$ \citep{2018LoureiroBOSS}. Covariances are calculated using 6000 log-normal mocks with {FLASK} \citep{Flask2016} and a spline to the data's $C_{\ell}$'s (to avoid introducing cosmological model assumptions). Because of the nature of the PCL estimator and partial sky observations, we forward model the mask effects into the likelihood, convolving the theory with the mixing matrix, $S_{\ell} = \sum_{\ell'}R_{\ell \ell'} C_{\ell'}$. Other effects such as redshift space distortions, shell crossing due to fingers of god, and extra Poissonian shot noise are incorporated through the theoretical auto- and cross-angular power spectra calculation. Detailed aspects related to the BOSS $C_{\ell}$ data vector, covariance matrix estimation, pipeline testing, and the implemented likelihood are outlined in detail in a previous paper \cite{2018LoureiroBOSS}.

We combine our BOSS angular power spectra with external data from the cosmic microwave background, supernovae type Ia (SNe Ia), and big bang nucleosynthesis (BBN) at the likelihood level using the \textit{unified cosmological library for parameter inference} code, or {UCLPI} \footnote{Cuceu \textit{et al.}, {(to be published)}. More details about the code and tests can be found in Ref. \cite{2018LoureiroBOSS}.}, which uses the primordial power spectra and transfer function from {CLASS} \cite{Class}. The CMB data used were the 2015 Planck CMB temperature, polarization and lensing measurements \citep{PlanckLikelihood2015}. The Planck likelihood uses low-$\ell$ modes for temperature (TT) and polarization auto- and cross-correlations (BB, TB, and EB). For higher multipoles $\ell > 30$, we used temperature (TT) and polarization auto- and cross-correlations (TE and EE)--a configuration known as Planck TT,TE,EE+lowTEB \citep{PlanckLikelihood2015,PlanckResults2015}. We also added the lensing likelihood based on both temperature and polarization maps. Next, we used  the most recent combined Pantheon SNe Ia sample \citep{2018Pantheon}. This sample contains 1048 SNe Ia in a redshift range $0.01 < z < 2.3$ and contains data from Pan-STARRS, SDSS, SNLS, and HST. The BBN information used in this work comes from measurements of the deuterium-hydrogen fraction estimated with recent improved helium-4 predictions as presented in Ref. \citep{2018BBN-Measurements}. The BBN likelihood was implemented with the help of the {AlterBBN} code \cite{2018AlterBBN}. We verified that the addition of BBN data does not have a direct impact on the neutrino mass parameters. BBN data help to constrain $N_{\text{eff}}$; better constraints on this parameter could have been achieved using extra BBN data such as He-4 (however, this is beyond the scope of this Letter).
\begin{figure}
\begin{center}
\includegraphics[width=\columnwidth]{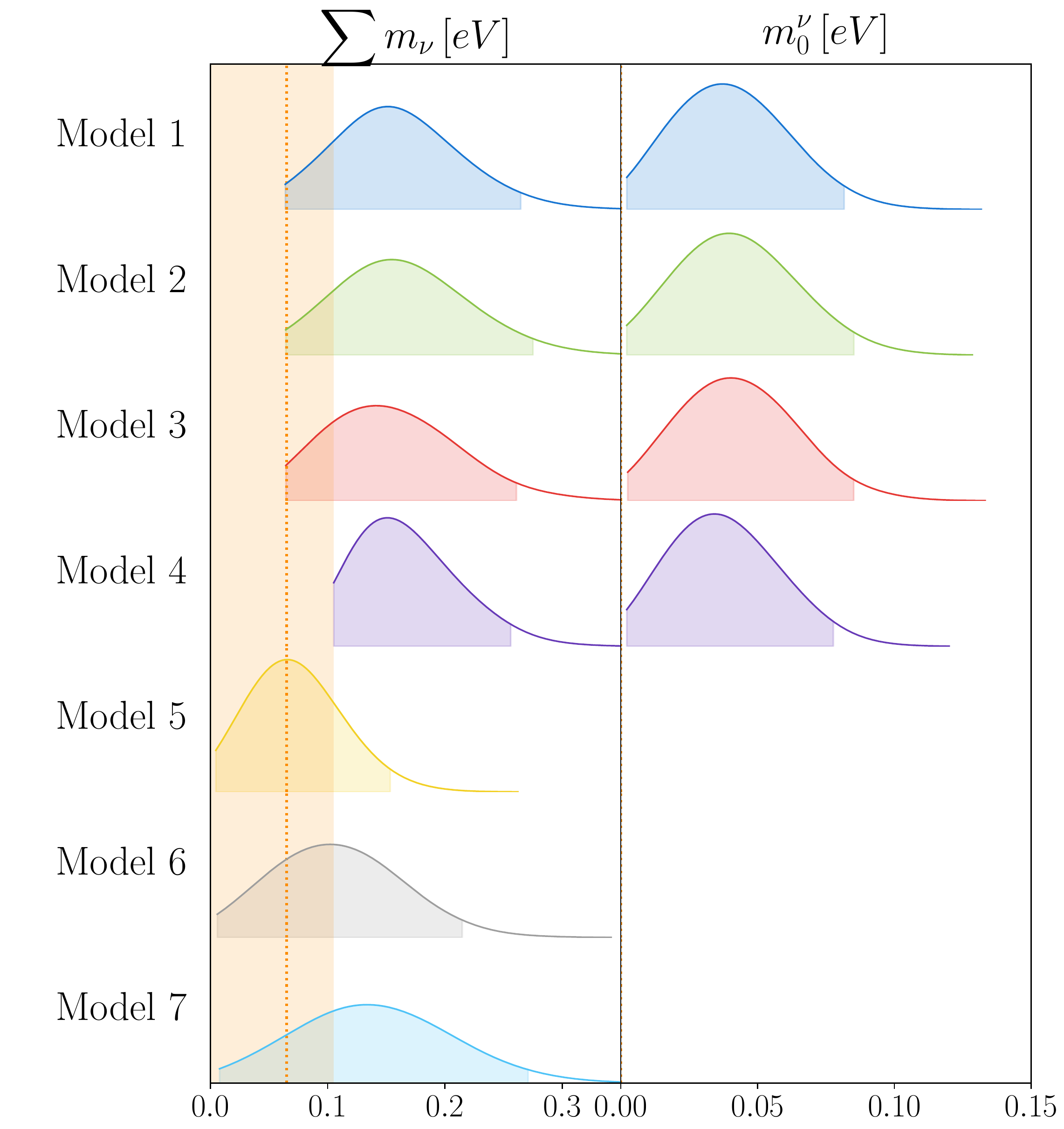}
\caption{The marginalized posterior probabilities for neutrino-related parameters for a range of neutrino models; the colored areas under the curves delineate the 95\% CI. Exact models (models 1-4) yield  robust constraints for the upper bound of $\sum m_{\nu} \lesssim 0.26$ eV (95\% CI) and for the lightest neutrino mass $m_0^{\nu} \lesssim 0.086$ eV (95\% CI), while models with commonly used cosmological approximations (models 5-7) have up to 43\% variation for the upper bound of $\sum m_{\nu}$ at 2$\sigma$ CI. The vertical dashed line in the left plot shows the minimum possible value for $\sum m_{\nu}$ for the NH, while the shaded region shows the same for the IH. The former region is excluded by particle physics experiments. All models also sample the basic $\Lambda$CDM parameters plus $N_{\text{eff}}$, shown in Fig. \ref{fig:LCDM}.} 
\label{fig:neutrinoCompare1}
\end{center}
\end{figure}

\textit{Analysis.}--We implemented nine different models to assess the impact of prior models on the upper bound of \NM{}. All models sample the basic $\Lambda$CDM parameters: $\{\Omega_b, \Omega_{cdm}, \ln 10^{10}A_s, n_s, h, \tau_{\text{reio}}\}$ as well as $N_{ur}$ to account for extra effective ultrarelativistic species.  All models were analyzed by varying $N_{\text{eff}}$ (directly or as a derived parameter) and therefore present a wider, stronger statement than would have been the case for a fixed value of $N_{\text{eff}}$. The posterior distribution analysis also contains several nuisance parameters for each of the datasets; these account for linear galaxy bias $b(z)$ and redshift dispersion $\sigma_s(z)$ for each of the 13 redshift tomographic bins in the BOSS dataset, two extra shot-noise parameters, $\mathcal{N}_{11}$ and  $\mathcal{N}_{12}$, for the last two bins in the BOSS data set due to the lower number of galaxies in each of them, the absolute SNe Ia magnitude in the \textit{B} band for the Pantheon sample, $M_B^{\text{PNT}}$, and the overall Planck calibration nuisance parameter, $y_{\text{cal}}^{\text{Planck}}$. These result in a total of 30 nuisance parameters, all marginalized over after the posterior is sampled. We performed the analysis using three different nested samplers: {Multinest} \cite{2009Multinest}, {Polychord} \cite{2015Polychord}, and {Pliny} \cite{PlinyRichardThesis}. The presented results are those from {Pliny}; the other samplers produced results that were essentially identical. Priors for the basic $\Lambda$CDM and nuisance parameters for this study are kept the same as in Table 3 in Ref. \citep{2018LoureiroBOSS}, a paper complementary to this work.

\begin{figure}
\begin{center}
\includegraphics[width=\columnwidth]{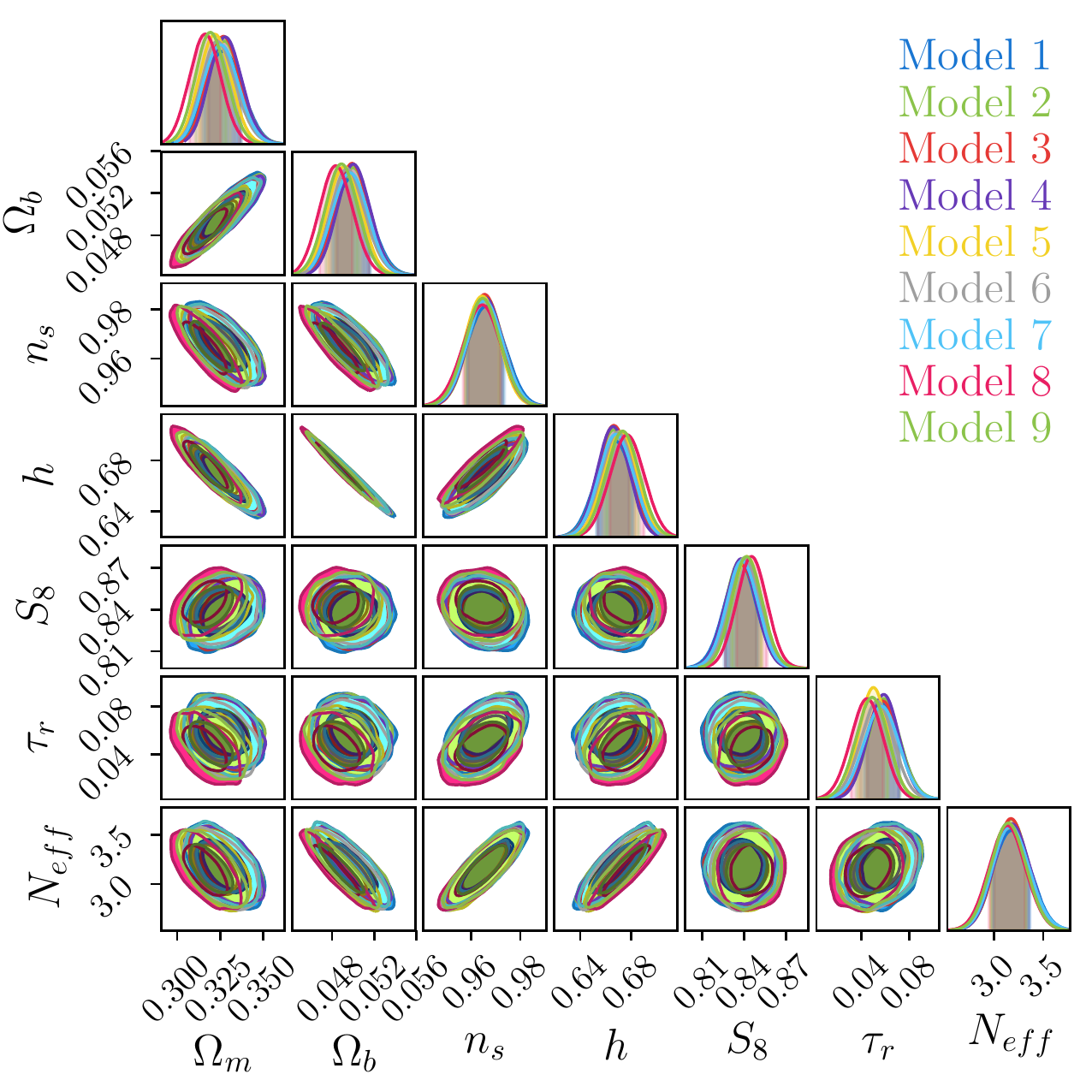}
\caption{One- (68\% CI) and two-dimensional (68\% and 95\% CI) marginalized posterior distributions for the relevant sampled and derived $\Lambda$CDM parameters considered in each of the nine different models (where $S_8 \equiv \sigma_8\sqrt{\Omega_m/0.3})$. All models agree in the basic $\Lambda$CDM parameters and for $N_{\text{eff}}$ to within half-$\sigma$ or less; model 8 is an outlier among the models, since it contains no massive neutrinos, and hence often yields a mild outlier among the marginalized posterior distributions. These results address the issue of how the modeling of neutrinos should be done within a standard $\Lambda$CDM analysis where the \NM{} is not the main focus of the analysis. It is clear that the simpler approach, leading to no biases, is the one taken by model 9 (the same as in Refs. \cite{PlanckCosmology2018,2016PlanckCosmology}). }
\label{fig:LCDM}
\end{center}
\end{figure}

We performed a full cosmological analysis for all models. The one-dimensional marginalized posteriors for \NM{} and the lightest neutrino mass $m_{0}^{\nu}$ can be found in Fig. \ref{fig:neutrinoCompare1}, while the upper bounds can be found in Table \ref{Tb:Models}. The standard $\Lambda$CDM parameters, together with $N_{\text{eff}}$, are shown in Fig. \ref{fig:LCDM}. This shows that all models essentially agree with each other, with a very small ($< 0.5 \sigma$) difference appearing only for the model with no massive neutrinos, model 8.

The marginalized posteriors for \NM{} (Fig. \ref{fig:neutrinoCompare1}) show that the use of exact models yields robust upper bounds at 95\% CI, varying between $< 0.256$ eV and $<0.275$ eV. The models in which the hierarchy was also sampled, models 1 and 2, did not demonstrate a significant choice between NH and IH; therefore, we marginalized over the hierarchy to get the results shown in Fig. \ref{fig:neutrinoCompare1}. Meanwhile, the commonly used cosmological approximations demonstrate a variation in the 95\% CI upper bound of 43\% between models 5 and 7--$\sum m_{\nu} < 0.154$ eV and $\sum m_{\nu} < 0.270$ eV, respectively. This indicates that approximations can be problematic since the upper bounds obtained are dominated by the prior model choice.

The ratio of evidences between two models, known as the Bayes factor, quantifies statistically if either is more strongly supported by the data \cite{1995BayesFactor}. The Bayes factors for all pairs of models considered were consistent with unity (to within the statistical precision of the nested sampling algorithm), meaning that the data used in this analysis do not strongly support any one of the models over the others.

\textit{Conclusions.}--We have shown that the choice of how the neutrino is modeled for cosmological purposes significantly affects current upper bounds for the sum of the neutrino masses. If physically motivated (exact) models are chosen, the upper bound is found to be $\sum m_{\nu} < 0.264$ eV (95\% CI). On the other hand, we now possess enough cosmological data to show that this upper bound is significantly different if we make the approximation that one (two) of the neutrino mass eigenstates have zero mass and that the mass is contained in the other two (one) eigenstates. 

We show here a concise framework, applied to the largest spectroscopic galaxy survey to date, to obtain robust neutrino mass information from a combination of cosmological observations and particle physics constraints. Even though no model was preferred from a Bayesian evidence analysis, cosmological approximations can cause a variation up to 43\% on the upper bound of \NM{}, while all exact models yield results that vary only by 7\% for the upper bound (both considered at 95\% CI). Using this exact modeling methodology, we present one of the first cosmological measurements of the upper bound of the lightest neutrino mass species: $m_{0}^{\nu} < 0.086$ eV at 95\% CI. 
Even though the posterior distributions for $m_{0}^{\nu}$ in Fig. \ref{fig:neutrinoCompare1} exhibit a peak, we do not claim it to be a detection (as the lower bound of the prior is not excluded by the 95\% CI).

In light of these results, we argue that the approach presented here as model 1 should be the choice for current and future cosmological neutrino mass investigations (given the volume of data now available to cosmologists). Even though the data used in this work still yield results within the degenerate mass spectrum scenario, we reinforce the idea that one should no longer make approximations, such as models 5 and 6, as these could lead to potentially nonphysical upper bounds and constraints. Instead, one should make use of a cosmological analysis that takes into account both of the neutrino mass hierarchies, as well as particle physics constraints and their uncertainties. 

Finally, we demonstrate that if neutrino masses are not the interest of the analysis, model 9 yields reliable cosmological results in the $\Lambda$CDM model context. In other words, a standard $\Lambda$CDM analysis is independent of the fiducial choice for the neutrino mass model, allowing for a simple approach to be taken. Following, note that changing the neutrino mass modeling does not affect $N_{\text{eff}}$. This suggests that if one wishes to study $N_{\text{eff}}$, the particular model chosen for the neutrino masses does not seem to play a role (see Fig. \ref{fig:LCDM}). We emphasize that one should consider massive neutrinos for a standard $\Lambda$CDM analysis--as the data are sensitive enough, as seen in the difference between the model with zero massive neutrinos (model 8) and all others in Fig. \ref{fig:LCDM}. The exact approach for neutrino mass estimation will be extremely relevant for future cosmological neutrino studies in the analysis of the next generation of surveys, e. g. DESI \citep{2016-DESI}, Euclid \citep{2011EuclidRedPaper}, LSST \citep{LSST}, and J-PAS \citep{JPAS}.

All cosmological contour plots were generated using {ChainConsumer} \cite{ChainConsumer}.

\vspace{0.3cm}

A.L. and B.M. thank the Brazilian people for support through Science without Borders fellowships from the Conselho Nacional de Desenvolvimento Científico e Tecnológico (CNPq). A.C. acknowledges the Royal Astronomical Society for support via a summer bursary and the United Kingdom Science and Technology Facilities Council (STFC) via a postgraduate studentship. B.M. and F.B.A. acknowledge support from the European Community through the DEDALE grant (Contract No. 665044) within the H2020 Framework Program of the European Commission. FBA acknowledges the Royal Society for support via a Royal Society URF. O.L. acknowledges support from a European Research Council Advanced Grant No. FP7/291329 and support from the United Kingdom Science and Technology Research Council (STFC) Grant No. ST/M001334/1. M.M. acknowledges support from the European Union's Horizon 2020 research and innovation program under Marie Sklodowska-Curie Grant Agreement No. 6655919. H.X. acknowledges the financial support from the Brazilian funding agency FAPESP. We also thank Constance Mahony, Tarso Franarin, and Pablo Lemos for their very helpful comments during the development of this work.

\bibliography{bibliog}
\end{document}